\documentclass[twocolumn,aps,prd,amsmath,amssymb]{revtex4}

\usepackage{bm}
\usepackage{amsmath}
\usepackage{amssymb}
\usepackage{latexsym}
\usepackage{amsfonts}
\usepackage{epsfig}
\usepackage{psfrag}
\usepackage{graphicx}

\usepackage{bm}
\usepackage{amsmath}
\usepackage{amssymb}
\usepackage{latexsym}
\usepackage{amsfonts}
\usepackage{epsfig}
\usepackage{psfrag}
\usepackage{graphicx}

\newcommand{\ul}{\underline}

\newcommand{\na}{\mbox{\boldmath$\nabla$}}
\newcommand{\bea}{\begin{eqnarray}}
\newcommand{\ea}{\end{eqnarray}}
\newcommand{\eea}{\end{eqnarray}}
\newcommand{\ord}{{\cal O}}

\begin{document}

\title{``Exotic'' quantum effects in the laboratory?} 

\author{Ralf Sch\"utzhold} 

\affiliation{Institut f\"ur Theoretische Physik,
Technische Universit\"at Dresden, D-01062~Dresden, Germany}

%\label{firstpage}

%\maketitle

\begin{abstract}%{Analogue gravity, Hawking radiation, Unruh effect}
This Article provides a brief (non-exhaustive) review of some recent 
developments regarding the theoretical and possibly experimental 
study of ``exotic'' quantum effects in the laboratory with 
special emphasis on cosmological particle creation, 
Hawking radiation, and the Unruh effect. 
\end{abstract}

\maketitle

%%%%%%%%%%%%%%%%%%%%%%%%%%%%%%%%%%%%%%%%%%%%%%%%%%%%%%%%%%%%%%%%%%%%%%%%%%%%%%%
%%%%%%%%%%%%%%%%%%%%%%%%%%%%%%%%%%%%%%%%%%%%%%%%%%%%%%%%%%%%%%%%%%%%%%%%%%%%%%%
\section{Introduction}
%%%%%%%%%%%%%%%%%%%%%%%%%%%%%%%%%%%%%%%%%%%%%%%%%%%%%%%%%%%%%%%%%%%%%%%%%%%%%%%
%%%%%%%%%%%%%%%%%%%%%%%%%%%%%%%%%%%%%%%%%%%%%%%%%%%%%%%%%%%%%%%%%%%%%%%%%%%%%%%

The combination of quantum field theory with non-trivial classical
background configurations -- such as a gravitational field -- yields
many fascinating  effects, see, e.g., Birrell \& Davies (1982) and
Fulling (1989).  
A striking example is the amplification of omnipresent quantum vacuum 
fluctuations by the influence of the classical background
configuration and thereby their conversion into real particles.
This phenomenon may occur in an expanding 
universe\footnote{According to our understanding of the evolution of our 
Universe, traces of such an effect -- i.e., the amplification of quantum 
vacuum fluctuations during the epoch of cosmic inflation -- can still be 
observed today in the anisotropies of the cosmic microwave background 
radiation.} 
(cosmological particle creation) 
and lies at the heart of Hawking radiation 
(Hawking, 1974, 1975), 
which is related to the Unruh effect 
(Unruh, 1976) 
via the principle of equivalence.

In order to discuss these phenomena, it is useful to recall some of
the basic properties of the quantum effects under consideration:
Since the created particles stem from quantum (vacuum) fluctuations,  
they vanish in classical limit ``$\hbar\to0$''. 
The particles are always created in pairs (squeezed vacuum state) 
in an entangled state (which is an inherently non-classical feature).
In cosmological particle creation, the two particles of each pair have 
opposite spins and momenta -- whereas, for the Hawking and Unruh
effects, the two partners occur on either side of the horizon. 
The two-particle states are pure quantum states while averaging over 
one partner yields the thermal density matrix (entanglement entropy).  

%%%%%%%%%%%%%%%%%%%%%%%%%%%%%%%%%%%%%%%%%%%%%%%%%%%%%%%%%%%%%%%%%%%%%%%%%%%%%%%
%%%%%%%%%%%%%%%%%%%%%%%%%%%%%%%%%%%%%%%%%%%%%%%%%%%%%%%%%%%%%%%%%%%%%%%%%%%%%%%
\section{The underlying analogy}
%%%%%%%%%%%%%%%%%%%%%%%%%%%%%%%%%%%%%%%%%%%%%%%%%%%%%%%%%%%%%%%%%%%%%%%%%%%%%%%
%%%%%%%%%%%%%%%%%%%%%%%%%%%%%%%%%%%%%%%%%%%%%%%%%%%%%%%%%%%%%%%%%%%%%%%%%%%%%%%

Apart from their quantum nature, which implies a suppression of these 
effects in a typical laboratory scenario due to the smallness of $\hbar$, 
they are intrinsically relativistic phenomena and hence additionally hard 
to observe in view of the large value of the speed of light $c$. 
Therefore, it can be advantageous to consider the analogues of these effects 
in appropriate condensed-matter systems -- where the speed of light $c$ 
should be replaced by the propagation velocity of the quasi-particles under
consideration 
(e.g., phonons which travel with the speed of sound, see Unruh, 1981). 
The analogy can be made explicit by writing down the most general 
linearized $\ord(\phi^3)$ low-energy $\ord(\partial^3)$ effective action 
for scalar Goldstone-mode quasi-particles $\phi$, which can be cast 
into the following form 
\bea
\label{general}
{\mathcal L}_{\rm eff}
=
\frac12(\partial_\mu\phi)(\partial_\nu\phi)G^{\mu\nu}(\ul x)
+\ord(\phi^3)
+\ord(\partial^3)
\,,
\ea
where $G^{\mu\nu}(\ul x)$ denotes some tensor which encodes the
details of the underlying condensed-matter system. 
Via the replacement 
$G^{\mu\nu}\to g^{\mu\nu}_{\rm eff}\sqrt{-g_{\rm eff}}$, we see that 
the kinematic aspects (i.e., the equations of motion and the commutation 
relations etc.) of the quasi-particles $\phi$ are completely equivalent 
to a scalar field in a curved space-time described by the effective metric
$g^{\mu\nu}_{\rm eff}$. 

Now, how can we exploit this analogy and what can we learn by using it?
Of course, the first answer could be that the aforementioned quantum 
effects have merely been predicted, but not directly observed yet. 
Hence, the analogy (\ref{general}) facilitates an experimental verification 
of these phenomena (by means of their laboratory analogues) and a test of 
the assumptions/approximations entering their derivation. 
Furthermore, it allows us to study the corrections imposed on these 
effects due to the impact of interactions, the back-reaction of the quantum 
fluctuations onto the classical background, the influence of a finite 
temperature bath and decoherence etc. 
Finally, apart from the purely experimental point of view, we may use
the condensed-matter analogues as theoretical toy models for quantum
gravity  
which inspire new ideas for studying the influence of the microscopic 
structure on macroscopic phenomena, for example. 

The other way around, the analogy (\ref{general}) may help us to understand
non-equilibrium (quantum) phenomena in general condensed-matter systems 
in terms of the vast amount of geometrical tools and concepts developed 
within general relativity and to extract universal features.
(For the low-energy quasi-particle kinematics, we may forget about the 
microscopic details of the underlying condensed-matter system and derive 
everything from the effective metric $g^{\mu\nu}_{\rm eff}$, see,
e.g., Sch\"utzhold, 2008; Sch\"utzhold \& Uhlmann, 2005; 
Sch\"utzhold \& Unruh, 2007).  

%%%%%%%%%%%%%%%%%%%%%%%%%%%%%%%%%%%%%%%%%%%%%%%%%%%%%%%%%%%%%%%%%%%%%%%%%%%%%%%
%%%%%%%%%%%%%%%%%%%%%%%%%%%%%%%%%%%%%%%%%%%%%%%%%%%%%%%%%%%%%%%%%%%%%%%%%%%%%%%
\section{UV catastrophe}
%%%%%%%%%%%%%%%%%%%%%%%%%%%%%%%%%%%%%%%%%%%%%%%%%%%%%%%%%%%%%%%%%%%%%%%%%%%%%%%
%%%%%%%%%%%%%%%%%%%%%%%%%%%%%%%%%%%%%%%%%%%%%%%%%%%%%%%%%%%%%%%%%%%%%%%%%%%%%%%

Before turning to experiments, let us discuss an example where condensed 
matter serves as a toy model for theoretical studies. 
The Hawking effect describes the evaporation of black holes and results in 
the emission of thermal radiation with the temperature 
(Hawking, 1974, 1975)
\bea
T_{\rm Hawking}
=
\frac{1}{8\pi M}
\frac{\hbar\,c^3}{G_{\rm N}k_{\rm B}}
\,,
\ea
where $M$ is the mass of the black hole. 
Hawking's derivation of this effect is based on the semiclassical approach 
of quantum fields propagating in a classical background space-time and the 
assumption that this treatment is valid up to arbitrarily large energies. 
However, there are many reasons to expect that this semiclassical approach 
fails at high energies, such as the Planck scale.
For example, one could argue that is should be impossible to resolve spatial 
details smaller than the Planck length 
$\ell_{\rm P}=\sqrt{\hbar G_{\rm N}/c^3}\approx1.6\times10^{-35}$m,
because the necessary (Planckian) momentum and hence energy concentrated 
in such a small volume would collapse into a clack hole. 

Inspired by the analogy to condensed matter
(cf.~Jacobson, 1991), a simple model for ultra-high 
energy deviations from classical general relativity is a modification
of the dispersion relation $\omega=ck\to\omega=\omega(k)$ of the
propagating degrees of freedom (e.g., photons, gravitons). 
E.g., in analogy to special relativity, where the acceleration of a
massive particle up to the speed of light is prohibited by a diverging
energy $E=mc^2/\sqrt{1-v^2/c^2}$, one could speculate that quantum
gravity impedes a sub-Planckian spatial resolution in a similar manner
via $\omega=ck/\sqrt{1-\ell_{\rm P}^2k^2}$.  
However, calculating the Hawking radiation for such a dispersion relation
(Sch\"utzhold \& Unruh, to appear; see also Unruh \& Sch\"utzhold,
2005; Brout {\em et al.}, 1995; Corley, 1998), 
we find that we reproduce the thermal emission at low energies 
$\ell_{\rm P}k\ll1$, but, in addition, we obtain a large amount of 
radiation (UV catastrophe) at ultra-high energies 
$k=\ord(1/\ell_{\rm P})$.
Since black holes are supposed to exist for a macroscopic time, the  
dispersion relation $\omega=ck/\sqrt{1-\ell_{\rm P}^2k^2}$ 
should be excluded as a realistic model for 
ultra-high energy deviations from classical general relativity. 
One way to avoid this UV catastrophe is a sub-luminal dispersion relation 
$d\omega/dk<c$. 

%%%%%%%%%%%%%%%%%%%%%%%%%%%%%%%%%%%%%%%%%%%%%%%%%%%%%%%%%%%%%%%%%%%%%%%%%%%%%%%
%%%%%%%%%%%%%%%%%%%%%%%%%%%%%%%%%%%%%%%%%%%%%%%%%%%%%%%%%%%%%%%%%%%%%%%%%%%%%%%
\section{Trapped ions}\label{Trapped ions}
%%%%%%%%%%%%%%%%%%%%%%%%%%%%%%%%%%%%%%%%%%%%%%%%%%%%%%%%%%%%%%%%%%%%%%%%%%%%%%%
%%%%%%%%%%%%%%%%%%%%%%%%%%%%%%%%%%%%%%%%%%%%%%%%%%%%%%%%%%%%%%%%%%%%%%%%%%%%%%%

For an experimental verification, there are several possible systems.
Let us start with trapped ions
(see also Sch\"utzhold {\em et al.}, 2008b). 
In a strongly elongated trap, the (quantized) positions $Q_i(t)$ of the 
ions satisfy the equation of motion 
\bea
\label{eom-full}
\ddot Q_i+\Omega_{\rm ax}^2(t) Q_i
=
\Gamma\sum\limits_{j \neq i}
\frac{{\rm sign}(i-j)}{(Q_i-Q_j)^2}
\,,
\ea
where $\Gamma$ encodes the strength of the Coulomb repulsion. 
If the axial trapping frequency $\Omega_{\rm ax}(t)$ is time-dependent,
the chain of ions expands or contracts linearly, which can be described in 
terms of the scale parameter $b(t)$.
Linearization $\hat Q_i(t)=b(t)Q_i^0+\delta\hat Q_i$ around the classical 
solution $b(t)Q_i^0$ determined by the initial positions $Q_i^0$
yields the wave equation of the (longitudinal) phonon modes 
$\delta\hat Q_i$
\bea
\label{linearization}
\left(\frac{\partial^2}{\partial t^2}+\Omega_{\rm ax}^2(t)\right)
\delta\hat Q_i
=
\frac{1}{b^3(t)}
\sum\limits_{j}
M_{ij}\delta\hat Q_j
\,\to\,
\frac{1}{b^3(t)}\na^2\delta\hat Q
\,.
\ea
The time-independent matrix $M_{ij}$ is strongly localized around the 
diagonal and approaches the second spatial derivative $\na^2$ in the
continuum limit. 
Hence, the above equation is analogous to a scalar field within an 
expanding/contracting universe with $b(t)$ corresponding to the 
scale parameter of the universe and $\Omega_{\rm ax}^2(t)$ being
roughly analogous to the Rici curvature scalar. 

Consequently, we may simulate cosmological particle creation in an 
ion trap:
Starting in the ground state of the ion chain at rest and subsequently 
varying the axial trapping frequency $\Omega_{\rm ax}(t)$ entails the 
creation of pairs of phonons (where both partner particles occupy the
same vibrational mode).
The advantage of this set-up lies in the well-developed pumping, 
cooling, and detection techniques, which are quite advanced in view of the 
potential of ion traps for quantum computing.
As a result, the proposed experiment should be realizable with present-day
technology and is presently in preparation. 
Via the comparison of first and second side-band transitions, it is even 
possible to test the two-phonon nature of this quantum effect and thereby
to distinguish it from classical effects such as heating. 
As a major draw-back, one should mention the limited size (number of ions),
for which a coherent control can be maintained.
Therefore, the analogue of cosmological particle creation is most probably 
feasible with this set-up, but Hawking radiation not yet. 

%%%%%%%%%%%%%%%%%%%%%%%%%%%%%%%%%%%%%%%%%%%%%%%%%%%%%%%%%%%%%%%%%%%%%%%%%%%%%%%
%%%%%%%%%%%%%%%%%%%%%%%%%%%%%%%%%%%%%%%%%%%%%%%%%%%%%%%%%%%%%%%%%%%%%%%%%%%%%%%
\section{Bose-Einstein condensates}\label{Bose-Einstein condensates}
%%%%%%%%%%%%%%%%%%%%%%%%%%%%%%%%%%%%%%%%%%%%%%%%%%%%%%%%%%%%%%%%%%%%%%%%%%%%%%%
%%%%%%%%%%%%%%%%%%%%%%%%%%%%%%%%%%%%%%%%%%%%%%%%%%%%%%%%%%%%%%%%%%%%%%%%%%%%%%%

One option to overcome the size limitation of ion traps -- while
maintaining low temperatures and coherent control -- is a Bose-Einstein
condensate (see, e.g., Dalfovo {\em et al.}, 1999). 
At length scales much larger than the healing length $\xi$
(typically of order $\mu$m), their dynamics can well be described by the 
Bernoulli equation.
Thus, the analogue of a black hole can be realized by a stationary 
flow which exceeds the speed of sound (of order mm/s) at some point. 
This point corresponds to the black hole horizon and thus emits 
Hawking radiation with the temperature being determined by the 
velocity gradient at that position (Unruh, 1981)
\bea
T_{\rm Hawking}
=
\frac{\hbar}{2\pi\,k_{\rm B}}\,
\left|\frac{\partial}{\partial r}\left(v_0-c_{\rm s}\right)\right|
\,.
\ea
Inserting the above values, we obtain typical temperatures of several 
nano-Kelvin, which could also be within reach of present-day technology
(Garay {\em et al.}, 2000; S.~W\"uster \& C.~M.~Savage, 2007, 
  %``Limits to the analogue Hawking temperature in a Bose-Einstein
  %condensate,'' 
{\tt arXiv:cond-mat/0702045}). 

Apart from the possibility of achieving very low temperatures, the
main advantages of Bose-Einstein condensates are the following:
Nowadays, we can handle relatively large samples containing many 
healing lengths $\xi=\ord(\mu{\rm m})$ whose shape can be influenced 
by laser fields, which represent external potentials for the Bernoulli
flow.  
This allows us to generate various velocity profiles for the condensate 
flow -- including quasi-one-dimensional flow -- without the usual problems 
of supersonic flow velocities (exceeding the Landau critical velocity,
cf.~Unruh, 2002) induced by the friction with walls etc. 
The detection of single (or pairs of) phonons is not quite as advanced as
for trapped ions, but could also be done in principle via similar optical 
techniques (see Sch\"utzhold, 2006). 

Unfortunately, there is also a major draw-back: 
All the gaseous Bose-Einstein condensates realized in the laboratory 
are only meta-stable states -- the true ground state is a solid. 
The main decay channels of the gaseous state are inelastic three-body 
collisions. 
Such an event transforms three indistinguishable bosons from the cloud 
into an molecule plus a remaining boson which carries the excess 
energy/momentum -- and thereby all three of them are effectively extracted 
from the condensate.
Now, if all three of these bosons stem from the same macroscopically 
occupied single-particle wave-function of the condensate ($k=0$), 
three-body collisions would just slowly diminish the number of 
condensed bosons.
However, in the presence of inter-particle interactions (which are necessary
for the propagation of sound), the many-particle ground state does also 
contain a small population of the higher single-particle states $k>0$.
This small fraction is called quantum depletion since it is generated by 
the quantum fluctuations (plus the interaction). 
Thus, if one of the three bosons involved in the inelastic collision stems 
from the quantum depletion and is removed, this event causes a deviation from 
the many-particle ground state -- i.e., an excitation 
(Dziarmaga \& Sacha, 2003). 
Ergo, three-body collisions do also heat up the condensate, which might 
swamp the Hawking signal to be detected.
In order to avoid this problem (and other issues, such as the black hole 
laser instability, see Corley \& Jacobson, 1999), it is probably desirable
to employ a  very fast detection method. 

%%%%%%%%%%%%%%%%%%%%%%%%%%%%%%%%%%%%%%%%%%%%%%%%%%%%%%%%%%%%%%%%%%%%%%%%%%%%%%%
%%%%%%%%%%%%%%%%%%%%%%%%%%%%%%%%%%%%%%%%%%%%%%%%%%%%%%%%%%%%%%%%%%%%%%%%%%%%%%%
\section{Electrons in laser fields}
%%%%%%%%%%%%%%%%%%%%%%%%%%%%%%%%%%%%%%%%%%%%%%%%%%%%%%%%%%%%%%%%%%%%%%%%%%%%%%%
%%%%%%%%%%%%%%%%%%%%%%%%%%%%%%%%%%%%%%%%%%%%%%%%%%%%%%%%%%%%%%%%%%%%%%%%%%%%%%%

As the final example, let us consider the Unruh effect (Unruh, 1976).  
This effect is related (though not equivalent) to Hawking radiation and 
states that an observer undergoing a uniform acceleration $a$ experiences 
the Minkowski vacuum as a thermal state with the Unruh temperature 
\bea
\label{unruh}
T_{\rm Unruh}
=
\frac{\hbar}{2\pi k_{\rm B}c}\,a
\,.
\ea
The most direct way of measuring this effect would be to accelerate a 
detector strong enough and to measure its excitations -- which is, 
however, very hard to do. 
Thus, we focus on an indirect signature:
An accelerated electron would also ``see'' a thermal bath of photons and 
hence it could scatter a photon out of this thermal bath into another mode.
Translation of this scattering event into the inertial frame corresponds
to the emission of two entangled photons by the accelerated electron
(Unruh \& Wald, 1984), which is a pure quantum effect.
In order to achieve a significant signal, the acceleration $a$ and thus the 
electric field $E$ accelerating the electron should be very large.
Hence, we propose sending an ultra-relativistic electron beam into a
counter-propagating laser pulse of high intensity, such that the electric 
field felt by the electrons is boosted and thus strongly amplified.
The probability that one of these electrons emits an Unruh pair 
of two entangled photons can be estimated via 
(Sch\"utzhold {\em et al.}, 2006, 2008a)
\bea
\label{prob-unruh}
{\mathcal P}_{\rm Unruh}
=
\frac{\alpha_{\rm QED}^2}{(4\pi)^2}
\left[\frac{E}{E_S}\right]^2
\times
\ord\left(
\frac{\omega T}{30}
\right)
\ll1
\,,
\ea
where $\alpha_{\rm QED}\approx1/137$ is the QED fine-structure constant and 
$E_S=m^2/q$ denotes the Schwinger limit (Schwinger, 1951). 
The electric field $E$, frequency $\omega$, and length $T$ of the laser 
pulse are measured in the rest frame of the electrons.
If $N_e=10^9$ electrons with a boost factor of $\gamma=300$ hit an optical 
laser pulse with 100 cycles, whose (laboratory-frame) intensity is of order 
$10^{18}\,\rm W/cm^2$, we obtain around one Unruh event in hundred shots.  
Since the above values are well within reach of present day-technology, 
the generation of photons pairs should be feasible\footnote{The signal 
could even be much more pronounced in undulators due to the coherent 
amplification (constructive interference of many electrons, 
Sch\"utzhold {\em et al.}, 2008a).}. 
Their detection and the elimination of the background noise is a bit
involved, but the unique features of this quantum effect (correlated
photon pairs, distinguished phase-space region etc.) should facilitate
a measurement. 

In order to point out the main advantage of this proposal, one should
realize that it is (in contrast to the previous examples in  
\S\ref{Trapped ions} and \S\ref{Bose-Einstein condensates}) a {\em real}
relativistic effect, i.e., the speed of light has not been replaced by
the speed of sound.  
Again, the generation of the photon pairs is most probably feasible with 
present-day technology and their detection should be within reach.
The major draw-back lies in the fact that it is not a direct measurement of 
the (original) Unruh effect: the acceleration is non-uniform and we did not 
consider the internal excitations of an accelerated detector, but an 
accelerated scatterer (i.e., electron).

%%%%%%%%%%%%%%%%%%%%%%%%%%%%%%%%%%%%%%%%%%%%%%%%%%%%%%%%%%%%%%%%%%%%%%%%%%%%%%%
%%%%%%%%%%%%%%%%%%%%%%%%%%%%%%%%%%%%%%%%%%%%%%%%%%%%%%%%%%%%%%%%%%%%%%%%%%%%%%%
\acknowledgements
%%%%%%%%%%%%%%%%%%%%%%%%%%%%%%%%%%%%%%%%%%%%%%%%%%%%%%%%%%%%%%%%%%%%%%%%%%%%%%%
%%%%%%%%%%%%%%%%%%%%%%%%%%%%%%%%%%%%%%%%%%%%%%%%%%%%%%%%%%%%%%%%%%%%%%%%%%%%%%%

This work was supported by the Emmy-Noether Programme 
of the German Research Foundation (DFG) under grant \# SCHU~1557/1-2,3.

%%%%%%%%%%%%%%%%%%%%%%%%%%%%%%%%%%%%%%%%%%%%%%%%%%%%%%%%%%%%%%%%%%%%%%%%%%%%%%%
%%%%%%%%%%%%%%%%%%%%%%%%%%%%%%%%%%%%%%%%%%%%%%%%%%%%%%%%%%%%%%%%%%%%%%%%%%%%%%%
\section*{Appendix: Classical analogues}
%%%%%%%%%%%%%%%%%%%%%%%%%%%%%%%%%%%%%%%%%%%%%%%%%%%%%%%%%%%%%%%%%%%%%%%%%%%%%%%
%%%%%%%%%%%%%%%%%%%%%%%%%%%%%%%%%%%%%%%%%%%%%%%%%%%%%%%%%%%%%%%%%%%%%%%%%%%%%%%

It might be illuminating to compare the proposals discussed here to some 
recent experiments devoted to effects in the vicinity of horizon analogues 
in the laboratory.
These analogues were based on surface waves 
(cf.~Sch\"utzhold \& Unruh, 2002) 
in water 
(G.~Rousseaux {\em et al.}, 2007, {\tt arXiv:0711.4767}) 
or helium 
(Volovik, 2003, 2005)  
and on light pulses in an optical fibre 
(T.~G.~Philbin {\em et al.}, 2007, {\tt arXiv:0711.4796}).  
First, it should be mentioned that the effects measured in these
experiments were not generated by (the analogues of) black-hole
horizons, but by other phenomena such as white-hole horizons and/or
ergo-regions. 
The white-hole horizon is the point where the flow velocity drops below the 
propagation speed of the waves.
For surface waves, this point is also known as hydraulic jump 
(cf.~Volovik, 2005). 
A white-hole horizon does not emit Hawking radiation but entails other 
classical and quantum instabilities.
Second, these experiments could only detect properties of 
classical\footnote{One of these striking classical phenomena is 
super-radiance, i.e., the amplification of wave-packets in 
rotating black holes (cf.~Sch\"utzhold \& Unruh, 2002).}
wave propagation, i.e., none of the quantum features mentioned in the 
Introduction (e.g., entangled pairs).
Clearly, water waves are most certainly not appropriate for observing 
quantum phenomena -- for superfluid helium and the optical fibres, this 
remains an open question.
   
%%%%%%%%%%%%%%%%%%%%%%%%%%%%%%%%%%%%%%%%%%%%%%%%%%%%%%%%%%%%%%%%%%%%%%%%%%%%%%%
%%%%%%%%%%%%%%%%%%%%%%%%%%%%%%%%%%%%%%%%%%%%%%%%%%%%%%%%%%%%%%%%%%%%%%%%%%%%%%%
%\end{acknowledgements}
%%%%%%%%%%%%%%%%%%%%%%%%%%%%%%%%%%%%%%%%%%%%%%%%%%%%%%%%%%%%%%%%%%%%%%%%%%%%%%%
%%%%%%%%%%%%%%%%%%%%%%%%%%%%%%%%%%%%%%%%%%%%%%%%%%%%%%%%%%%%%%%%%%%%%%%%%%%%%%%
   
%%%%%%%%%%%%%%%%%%%%%%%%%%%%%%%%%%%%%%%%%%%%%%%%%%%%%%%%%%%%%%%%%%%%%%%%%%%%%%%
%%%%%%%%%%%%%%%%%%%%%%%%%%%%%%%%%%%%%%%%%%%%%%%%%%%%%%%%%%%%%%%%%%%%%%%%%%%%%%%
%%%%%%%%%%%%%%%%%%%%%%%%%%%%%%%%%%%%%%%%%%%%%%%%%%%%%%%%%%%%%%%%%%%%%%%%%%%%%%%
%%%%%%%%%%%%%%%%%%%%%%%%%%%%%%%%%%%%%%%%%%%%%%%%%%%%%%%%%%%%%%%%%%%%%%%%%%%%%%%
%\begin{thebibliography}
%%%%%%%%%%%%%%%%%%%%%%%%%%%%%%%%%%%%%%%%%%%%%%%%%%%%%%%%%%%%%%%%%%%%%%%%%%%%%%%
%%%%%%%%%%%%%%%%%%%%%%%%%%%%%%%%%%%%%%%%%%%%%%%%%%%%%%%%%%%%%%%%%%%%%%%%%%%%%%%
%%%%%%%%%%%%%%%%%%%%%%%%%%%%%%%%%%%%%%%%%%%%%%%%%%%%%%%%%%%%%%%%%%%%%%%%%%%%%%%
%%%%%%%%%%%%%%%%%%%%%%%%%%%%%%%%%%%%%%%%%%%%%%%%%%%%%%%%%%%%%%%%%%%%%%%%%%%%%%%

%%%%%%%%%%%%%%%%%%%%%%%%%%%%%%%%%%%%%%%%%%%%%%%%%%%%%%%%%%%%%%%%%%%%%%%%%%%%%%%
%%%%%%%%%%%%%%%%%%%%%%%%%%%%%%%%%%%%%%%%%%%%%%%%%%%%%%%%%%%%%%%%%%%%%%%%%%%%%%%
\section*{References}
%%%%%%%%%%%%%%%%%%%%%%%%%%%%%%%%%%%%%%%%%%%%%%%%%%%%%%%%%%%%%%%%%%%%%%%%%%%%%%%
%%%%%%%%%%%%%%%%%%%%%%%%%%%%%%%%%%%%%%%%%%%%%%%%%%%%%%%%%%%%%%%%%%%%%%%%%%%%%%%

\begin{itemize}

\item % 
Birrell, N.~D.~\& Davies, P.~C.~W.~1982 
{\em Quantum Fields in Curved Space},
Cambridge University Press, Cambridge, UK. 

\item %
Brout, R., Massar, S., Parentani, R.~\& Spindel, P.~1995
Hawking Radiation Without Transplanckian Frequencies.
{\em Phys.\ Rev.\ D} {\bf 52}, 4559.

\item % 
Corley, S.~1998
Computing the spectrum of black hole radiation in the presence
of high  frequency dispersion: An analytical approach.
{\em Phys.\ Rev.\ D} {\bf 57}, 6280.

\item % 
Corley, S.~\& Jacobson, T.~1999
Black hole lasers.
{\em Phys.\ Rev.\ D} {\bf 59}, 124011.

\item %
Dalfovo, F., Giorgini, S., Pitaevskii, L.~P.~\& Stringari, S.~1999
Theory of Bose-Einstein condensation in trapped gases.
{\em Rev.\ Mod.\ Phys.} {\bf 71}, 463.

\item % 
Dziarmaga, J.~\& Sacha, K.~2003
Bose-Einstein-condensate heating by atomic losses.
{\em Phys.\ Rev.\ A} {\bf 68}, 043607.

\item % 
Fulling, S.~A.~1989 
{\em Aspects of Quantum Field Theory in Curved Space-Time},
Cambridge University Press, Cambridge, UK. 

\item %  
Garay, L.~J., Anglin, J.~R., Cirac, J.~I.~\& Zoller, P.~2000 
Sonic Analog of Gravitational Black Holes in Bose-Einstein Condensates. 
{\em Phys.\ Rev.\ Lett.} {\bf 85}, 4643.

\item %
Hawking, S.~W.~1974 
Black Hole Explosions.
{\em Nature} {\bf 248}, 30. 

\item %
Hawking, S.~W.~1975 
Particle Creation by Black Holes.
{\em Commun.\ Math.\ Phys.} {\bf 43}, 199.

\item %
Jacobson, T.~1991
Black Hole Evaporation and Ultrashort Distances.
{\em Phys.\ Rev.\ D} {\bf 44}, 1731.

\item % 
Sch\"utzhold, R.~2006
On the detectability of quantum radiation in Bose-Einstein
condensates.
{\em Phys.\ Rev.\ Lett.} {\bf 97}, 190405. 

\item % 
Sch\"utzhold, R.~2008
Emergent Horizons in the Laboratory.
{\em Class.\ Quantum Grav.} in press. 

\item %
Sch\"utzhold, R., Schaller, G.~\& Habs, D.~2006 
Signatures of the Unruh effect from electrons accelerated by
ultra-strong laser fields.
{\em Phys.\ Rev.\ Lett.} {\bf 97}, 121302. 

\item %
Sch\"utzhold, R., Schaller, G.~\& Habs, D.~2008 
Table-top creation of entangled multi-keV photon pairs and the Unruh
effect. 
{\em Phys.\ Rev.\ Lett.} in press. 
 
\item %
Sch\"utzhold, R., Uhlmann, M., Petersen, L., Schmitz, H., 
Friedenauer, A.~\& Sch\"atz, T.~2008 
Analogue of cosmological particle creation in an ion trap.
{\em Phys.\ Rev.\ Lett.} in press. 

\item % 
Sch\"utzhold, R.~\& Uhlmann, M.~2005
{\em Horizon Analogues in the Laboratory},
Proceedings of the Memorial Symposium for Gerhard Soff, 
Frankfurt, Germany. 

\item % 
Sch\"utzhold, R.~\& Unruh, W.~G.~2002
Gravity wave analogues of black holes. 
{\em Phys.\ Rev.\ D} {\bf 66} 044019.

\item % 
Sch\"utzhold, R.~\& Unruh, W.~G.~(eds.) 2007
{\em Quantum Analogues: From Phase Transitions to Black Holes \&
  Cosmology}, 
Springer Lecture Notes in Physics {\bf 718}. 

\item % 
Schwinger, J.~1951
On Gauge Invariance and Vacuum Polarization.
{\em Phys.\ Rev.} {\bf 82}, 664.

\item % 
Unruh, W.~G.~1976
Notes on Black Hole Evaporation.
{\em Phys.\ Rev.\ D} {\bf 14}, 870.

\item % 
Unruh, W.~G.~1981
Experimental Black Hole Evaporation?
{\em Phys.\ Rev.\ Lett.} {\bf 46}, 1351. 

\item % 
Unruh, W.~G.~2002
Measurability of Dumb Hole Radiation?
In {\em Artificial Black Holes} 
(eds M.~Novello, M.~Visser, and G.~Volovik)
World Scientific, Singapore. 

\item % 
Unruh, W.~G.~\& Sch\"utzhold, R.~2005
On the universality of the Hawking effect.
{\em Phys.\ Rev.\ D} {\bf 71}, 024028. 

\item % 
Unruh, W.~G.~\& Wald, R.~M.~1984
What Happens When An Accelerating Observer Detects A Rindler Particle. 
{\em Phys.\ Rev.\ D} {\bf 29}, 1047. 

\item %
Volovik, G.~E.~2003 
{\em The Universe in a Helium Droplet}, 
Oxford University Press, Oxford, UK. 

\item %
Volovik, G.~E.~2005
The hydraulic jump as a white hole.
{\em Pisma Zh.\ Eksp.\ Teor.\ Fiz.} {\bf 82} 706
(English version {\em JETP Lett.} {\bf 82} 624). 

\end{itemize}

%%%%%%%%%%%%%%%%%%%%%%%%%%%%%%%%%%%%%%%%%%%%%%%%%%%%%%%%%%%%%%%%%%%%%%%%%%%%%%%
%%%%%%%%%%%%%%%%%%%%%%%%%%%%%%%%%%%%%%%%%%%%%%%%%%%%%%%%%%%%%%%%%%%%%%%%%%%%%%%
%%%%%%%%%%%%%%%%%%%%%%%%%%%%%%%%%%%%%%%%%%%%%%%%%%%%%%%%%%%%%%%%%%%%%%%%%%%%%%%
%%%%%%%%%%%%%%%%%%%%%%%%%%%%%%%%%%%%%%%%%%%%%%%%%%%%%%%%%%%%%%%%%%%%%%%%%%%%%%%
%\end{thebibliography}
%%%%%%%%%%%%%%%%%%%%%%%%%%%%%%%%%%%%%%%%%%%%%%%%%%%%%%%%%%%%%%%%%%%%%%%%%%%%%%%
%%%%%%%%%%%%%%%%%%%%%%%%%%%%%%%%%%%%%%%%%%%%%%%%%%%%%%%%%%%%%%%%%%%%%%%%%%%%%%%
%%%%%%%%%%%%%%%%%%%%%%%%%%%%%%%%%%%%%%%%%%%%%%%%%%%%%%%%%%%%%%%%%%%%%%%%%%%%%%%
%%%%%%%%%%%%%%%%%%%%%%%%%%%%%%%%%%%%%%%%%%%%%%%%%%%%%%%%%%%%%%%%%%%%%%%%%%%%%%%
   
%\label{lastpage}
   
\end{document}